\documentclass[usenatbib]{mn2e}
\usepackage{times,graphicx,url}

\newcommand{\be}{\begin{equation}}
\newcommand{\ee}{\end{equation}}
\newcommand{\iraf}{{\sc IRAF}}

\title[Difference imaging with bright variable backgrounds]{Difference image
photometry with bright variable backgrounds}

\author[E.~Kerins et al.]{E.~Kerins,$^1$ M.J.~Darnley,$^2$ 
J.P. Duke,$^2$ A. Gould,$^3$
C. Han,$^4$ A.
Newsam,$^2$ 
\newauthor B.-G.
Park$^5$ and R.~Street$^6$\\
$^1$Jodrell Bank Centre for Astrophysics, School of Physics \&
Astronomy, University of Manchester, Oxford Road, Manchester M13 9PL\\
$^2$Astrophysics Research Institute, Liverpool John
Moores University, Twelve Quays House, Egerton Wharf, Birkenhead, Merseyside
CH41
1LD\\
$^3$Department of Astronomy, Ohio
State University, 140 West 18th Avenue, Columbus, OH 43210, USA\\
$^4$Department of Physics, Chungbuk National University,
Chongju 361-763, Korea\\
$^5$Korea
Astronomy and Space Science Institute, Hwaam-Dong, Yuseong-Gu, Daejeon
305-348, Korea\\
$^6$Las Cumbres Observatory Global Telescope Network, 6740 Cortona Drive, Suite
102, Goleta, CA 93117, USA
}

\begin{document}

\maketitle

\begin{abstract}
Over the last two decades the Andromeda Galaxy (M31) has been something of a
test-bed for methods aimed at obtaining accurate time-domain relative photometry
within highly crowded fields. Difference imaging methods, originally
pioneered towards M31, have evolved into sophisticated methods, such as
the Optimal Image Subtraction (OIS) method of \cite{ala98}, that today are
most widely used to survey variable stars, transients and microlensing events in
our own Galaxy. We show that modern difference image (DIA) algorithms such as
OIS, whilst spectacularly successful towards the Milky Way bulge, may perform
badly towards high surface brightness targets such as the M31 bulge. Poor
results can occur in the presence of common
systematics which add spurious flux contributions to images, such as internal
reflections, scattered light or fringing. Using data
from the Angstrom
Project microlensing survey of the M31 bulge, we show that very good results are
usually obtainable by first performing careful photometric
alignment prior to using OIS to perform point-spread function (PSF) matching.
This separation of background matching and PSF matching, a common feature of
earlier M31 photometry techniques, allows us to take full advantage of the
powerful PSF matching flexibility offered by OIS towards high surface
brightness targets. We find that difference images
produced this way have noise distributions close to Gaussian, showing
significant
improvement upon results achieved using OIS alone. We show that with this
correction
light-curves of variable stars and transients can be recovered to within $\sim
10$ arcseconds of the M31 nucleus. Our method is simple to implement and is
quick enough to be incorporated within real-time DIA pipelines. We also
demonstrate that OIS is remarkably robust even when, as in the case of the
central regions of the M31 bulge, the sky density of variable sources approaches
the confusion limit.
\end{abstract}


\section{Introduction}

Difference Image Analysis (DIA) is now used routinely to provide very accurate
relative photometry of variable stars, transient objects and microlensing
events in the Milky Way and other nearby galaxies \citep{woz08}. DIA permits
very accurate relative photometry even within extremely dense stellar fields
where conventional photometric methods can fail or suffer from serious bias.

Most DIA pipelines currently in use derive from the Optimal Image Subtraction
(OIS) algorithm of \cite{ala98} and \cite{ala00}. The OIS algorithm has
found widest use in providing accurate relative photometry of stars in the Milky
Way and the Magellanic Clouds. Similar schemes that pre-date
OIS were originally employed to look for microlensing towards the Andromeda
Galaxy
\citep{tom96,ans97}. Since the stars in the Andromeda Galaxy (M31) are two
orders of
magnitude further away than those typically monitored in our Galaxy, difference
imaging towards M31 throws up several additional challenges to the standard
technique. 

In this paper we show that towards the bulge of M31, and similarly towards other
targets where diffuse background surface brightness dominates
the total flux, DIA pipelines based on the OIS algorithm can often yield
poor results due to common image systematics such as internal
reflections, scattered light or fringe effects. Systematics,
which may appear at a low level
($\sim 1\%$) on the original
exposures, can give rise to large-amplitude differential background residuals
on difference image frames. Since OIS minimizes mismatches in both the point
spread function (PSF) and the differential background simultaneously, poor
background matching often results in poor PSF matching and therefore
substantial systematic errors in differential photometry.
We propose a straightforward remedy to allow the effects of such systematics
to be minimized. 

The structure of the paper is as follows. In Section~\ref{dialimits} we
briefly describe how M31 has been used as a test-bed in developing
time-domain photometry towards crowded stellar fields. The evolution of these
methods culminated in the Optimal Image Subtraction (OIS) algorithm
of \cite{ala98} and \cite{ala00}, which forms the basis for most DIA pipelines
currently in use.  We use images obtained by the Angstrom
Project \citep{ker06} of the bulge region of M31 to
show how the OIS algorithm may not perform optimally in the presence of
bright backgrounds. In
Section~\ref{moddia} we show how difference images with noise levels close to
the photon noise limit can be
recovered by separating the photometric alignment and
PSF matching stages. In Section~\ref{examplelc} we show some example
periodic variable light-curves from the Angstrom Project dataset, illustrating
the impact of the correction on their photometric quality and on
the ability to characterise variable stars at a range of distances from the
M31 core. We discuss the findings in
Section~\ref{discuss}.

\section{Difference imaging towards the
M31 bulge} \label{dialimits}

\subsection{Relative photometry in very crowded fields}

The Andromeda Galaxy has provided something of a test-bed for the development of
algorithms designed to obtain accurate relative photometry within crowded
fields. At around the same time two techniques emerged, difference imaging
\citep{tom96} and super-pixel photometry \citep{ans97}. Both techniques deal
with
the difficulty of obtaining robust relative photometry across epochs in the
presence of variations in seeing and sky background. 

\cite{tom96} advocated
the difference imaging approach, in which they convolve a high quality reference
image with a Gaussian kernel to match the point spread function
(PSF) of a target image. Prior to the convolution step the smooth M31 background
light was subtracted from both images using a local median
filter. The PSF matching approach
forms
the basis of most current difference imaging codes, though it assumes the
basic functional form of the PSF is known a priori. Modern
implementations do not include a separate background subtraction stage but
instead solve for the PSF and background simultaneously.

In the super-pixels method \citep{ans97}
an image pair is also photometrically aligned in order to match the background
levels. This was done first through linear photometric alignment and then by
local median filtering the images to produce smooth background maps. The
background map was subtracted from the target image and replaced with
that of the reference image.  In the absence of seeing
variations, or intrinsic source variability, the distribution of image
flux deviations (measured within square pixel arrays, or super-pixels)
with respect to the local background flux should be statistically the same
between aligned image pairs. By plotting the two distributions against one
another and determining the best linear fit, a simple linear flux correction can
be made to the target image flux to correct for seeing changes.
The super-pixels technique has the advantage of being empirically calibrated,
avoiding strong assumptions on the form of the PSF. However in
principle it is not as sensitive as difference imaging due to the potential loss
of signal through binning flux into super-pixels.

Interestingly both of these methods separate out the background correction from
the seeing correction. Modern difference imaging
algorithms make these two corrections simultaneously. Whilst this is a highly
efficient approach, as we shall show, it does not always yield optimal results.

\subsection{Optimal image subtraction}

Most difference image packages currently in use employ the Optimal Image
Subtraction (OIS) algorithm presented by \cite{ala98} and generalized further by
\cite{ala00}. In the OIS method a reference image $R$ is convolved to match the
PSF of a target image $T$ that has been geometrically
aligned and re-sampled to the same pixel grid as $R$. By taking into account
differences in the background between $R$ and $T$ via a smooth 2D background
model, $B$, the difference image, $D$, can be computed from linear least-squares
minimization:
\begin{eqnarray}
   \sum_i D(x_i,y_i)^2 & = & \min \{ \sum_i [(R(x_i,y_i) \otimes K(u,v)) -
\nonumber \\
& & T(x_i,y_i)+B(x_i,y_i)]^2/\sigma(x_i,y_i)^2 \}
\label{dia}
\end{eqnarray}
where $(x,y)$ denote the position of image pixel $i$, $K$ is a kernel function
that describes the PSF transformation between $R$ and $T$, $(u,v)$ are
coordinates centred on the kernel, $\sigma^2$ is the pixel variance and
$\otimes$ denotes convolution. For
computational
efficiency \cite{ala98} advocate decomposing $K$ into a linear combination of
basis functions
\be
   K(u,v)=\sum_{k=1}^N \sum_{p,q=0}^{p, p+q\leq M_k} a_{pq} u^p v^q
e^{-(u^2+v^2)/2\sigma_k^2}, \label{basis}
\ee
where $a_{pq}$ are coefficients. The Gaussian widths $\sigma_k$ and integers
$N$ and $M_k$ control the complexity of the kernel shape. Similarly the
differential background is modelled as a sum of polynomial basis functions:
   \begin{equation}
 B(x,y) = \sum_{r,s=0}^{r,r+s\leq M_b} a_{rs} x^r y^s, \label{bgbasis}
\end{equation}
where $a_{rs}$ are coefficients and the integer $M_b$ is the degree of
polynomial used to model the differential background.

There are a number of freely available software implementations of the OIS
algorithm, the best known of which are
ISIS\footnote{\url{http://www.iap.fr/users/alard/package.html}} and
DIAPL\footnote{\url{http://users.camk.edu.pl/pych/DIAPL/}}. Throughout
this paper we use ISIS version 2.2 \citep{ala98,ala00}. However, we stress
that the shortcomings we illustrate are not specific to ISIS but are shared by
all implementations of OIS that assume simple polynomial forms for the
differential background as in Equation~(\ref{bgbasis}).
As pointed out in the review of \cite{woz08} the OIS method is
actually independent of the choice of basis function. \cite{bram08} has
implemented a version of OIS (DANDIA) in
which
Equation~(\ref{dia}) is minimized pixel-by-pixel (i.e. essentially using a
$\delta$-function kernel). This method solves for the kernel for an assumed
constant background within a sub region. The global differential
background solution is then found via interpolation over a grid of
sub-regions. However for complex differential backgrounds of the kind
investigated in this paper the method would potentially require splitting the
image into a very large number of sub-regions in the absence of prior knowledge
on
the shape of the differential background and its behaviour with respect to time.

\begin{figure*}
\includegraphics[width=0.9\textwidth]{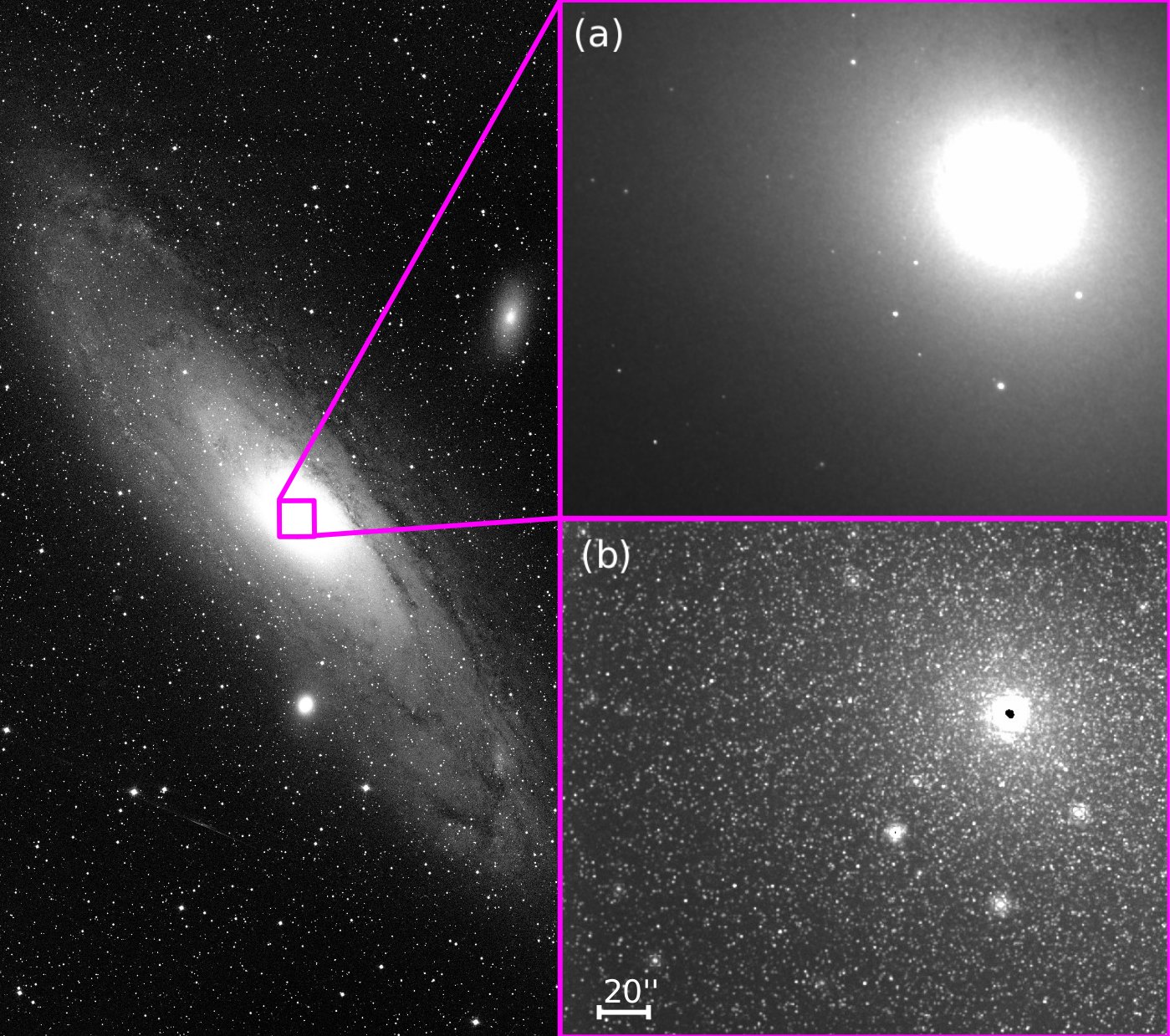}
\caption{{\em Left panel}\/: The position of the Angstrom Project robotic field
shown on an NOAO image of the Andromeda Galaxy (M31 - image credit: Bill
Schoening, Vanessa Harvey/REU program/NOAO/AURA/NSF). {\em Right panels}\/: (a)
A Sloan $i'$-band stacked image of the Angstrom Project robotic field obtained
from the Liverpool Telescope at La~Palma. This image is used as our reference
image for the tests in this paper and covers the inner $\sim 4' \times 4'$ of
the M31 bulge. (b) The locations of thousands of variable stars around the M31
core are revealed by combining a stack of 480 difference images. The difference
images span 5 years and are created from a sequence of more than 3300 individual
exposures. The difference images were created using the modified method
described in this paper. Composite residuals from imperfect subtraction within
the inner $\sim 10''$ of the core and around some bright foreground stars are
also visible.}
\label{angfld}
\end{figure*}

\subsection{Difference imaging of Angstrom Project data}

We have applied the OIS method to a series of Sloan $i'$-band images of the M31
bulge
obtained by the Angstrom Project \citep{ker06} using the Liverpool Telescope
\citep[hereafter LT,][]{ste04} on the Canary Islands. The Angstrom Project is
undertaking long-term
monitoring of the bulge of M31 in order to detect stellar microlensing events
and transients. The survey employs a real-time difference imaging pipeline
capable of issuing alerts of ongoing events \citep{dar07}. The location of the
Angstrom
robotic telescope field is shown in Figure~\ref{angfld}.

The LT is a 2m robotic telescope housing an optical camera (RATCAM) with a
$4\farcm 6 \times 4\farcm 6$ field of view with $0\farcs13$ pixels. Images  were
obtained over a 5-year period from 2004-2009. The LT typically obtained one or
two epochs of data per clear night during the M31 observing season
(August-January) with each epoch comprising a run of between 7 and 15 short
exposures, typically ranging from 30 to 200 seconds. The exposures were kept
short to avoid saturation of the M31 core and bright foreground stars.
Pre-processing, comprising de-biasing and flat fielding, was carried out by the
LT robotic telescope pipeline. The Angstrom Project pipeline then performed
preliminary geometric alignment and de-fringing, followed by optimal image
alignment using Fourier cross correlation \cite[for further details
see][]{dar07}. For the real-time pipeline the images are re-binned to
$1024\times 1024$ pixels with a resolution of $0\farcs 26$. In this paper we
also use these re-binned images.

All images were built by aligning and stacking individual exposures within each
observing epoch to form a high signal-to-noise epoch frame.  The master
reference frame was constructed from a stack of 30
high quality individual exposures obtained over four separate epochs from the
2007/8 and 2008/9 observing seasons.
These epochs were chosen on the basis of accurate telescope pointing and
tracking, low airmass, low background levels and good seeing. 
ISIS can itself be used to build the reference image from a stack of high
quality frames by matching their PSFs and background using
Equation~(\ref{dia}). We choose instead to create our reference frame by
first photometrically aligning the selected images (using the method
described in Section~\ref{photscale}) and then median stacking them using the
\iraf\footnote{\iraf\ is distributed by the
National Optical Astronomy Observatories,
which are operated by the Association of Universities for Research
in Astronomy, Inc., under cooperative agreement with the National
Science Foundation.} {\sc imcombine} task. Hence our difference image results
rest only on a
single application of Equation~(\ref{dia}) when matching the reference and
target images. The same reference image is used throughout so
that variations in our results depend only on how we manipulate the target
frame. The reference frame is shown in panel~(a) of Figure~\ref{angfld}.

\begin{figure*}
\begin{center}
\includegraphics[width=1.\textwidth]{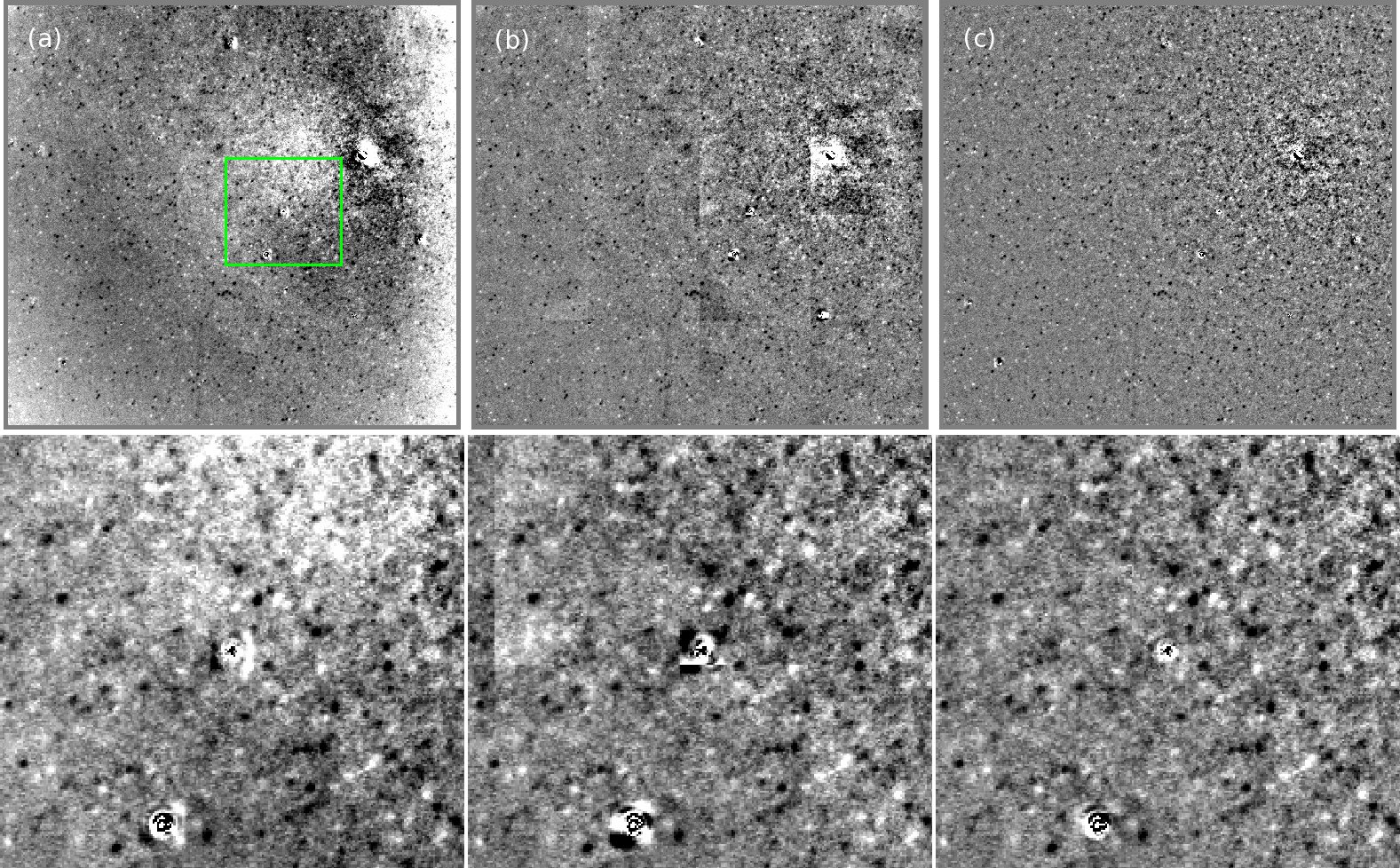}
\end{center}
\caption{{\em Upper panels}\/: difference images produced by ISIS (version 2.2)
from a pair of high
quality
reference and target images covering an area of
$3\farcm 9 \times 3\farcm 7$ around the M31 core. In (a) ISIS is
run on
photometrically unaligned images allowing a second-order polynomial to
model the differential background. Strong background residuals remain in the
difference image, which do not markedly improve with the use of higher order
polynomials.
In (b) the photometrically unaligned images are split into 16 sub-regions
on which ISIS is run individually using
the same parameters as in (a). The background matching is better further out
from the core (apparently at the expense of poorer PSF matching) but remains
poor
closer to the core where the boundaries between the sub-regions are evident.
In (c) the images have been photometrically aligned using the method
discussed in this paper, prior to
running ISIS. In this case a zeroth-order polynomial has been imposed for the
differential background when running ISIS. All other parameters are the same as
in (a) and (b). {\em Lower panels}\/: show a
$1' \times 0\farcm 9$ zoom of the region shown by the box in (a) for each case
(a), (b) and (c),
respectively. In (c) note how the removal
of the differential background
also improves the quality of PSF matching as evidenced by the lessened
residuals around bright foreground stars. 
}
\label{isis-bg}
\end{figure*}

Figure~\ref{isis-bg} shows some example difference
images created from the reference image and a very high quality
target image obtained under excellent seeing. The results illustrate the
challenge of accurate background matching towards the core of M31 where, even
within the small LT field of view, the dynamic range in surface brightness
across the image is around two orders of magnitude. In
Figure~\ref{isis-bg}(a) ISIS uses a second-order polynomial to model the
differential background across the whole image area. Whilst the resulting
difference image clearly succeeds in revealing a wealth of variable stars
(evidenced by the black and white spots, which show sources that have dimmed
or brightened) it
suffers from a significant large-scale background residual. We find that the
amplitude of the
residual is largely insensitive to the choice of polynomial order, though the
residual pattern across the image does vary according to the order used, as one
would expect. The underlying differential background is therefore clearly not
well represented by a 2D polynomial. There are a number of possible causes for
such an effect. Scattered light, internal reflections, dust and fringing
effects can add localized unwanted flux contributions, whilst flat
fielding errors may have a multiplicative effect. When observing a target like
the M31 bulge which has a very bright and spatially varying background, the
effects
of such systematics are unlikely to be adequately modelled by a polynomial
function.

In fact the amplitude of the background
variation over most of the image area in Figure~\ref{isis-bg}(a) typically
represents only $\la 1\%$ of the original image flux, but this is still too
large to
allow reliable difference image photometry of sources, which themselves are
varying at a level of a few percent of the background surface brightness flux.
Since the residual background is epoch
dependent the failure to effectively remove it means that robust relative
photometry between epochs becomes virtually impossible. Additionally, the
large-amplitude
background
residual may blunt the OIS algorithm's sensitivity
to the optimal PSF transformation because the overall summed squared difference
flux
[$D^2$ in Equation~(\ref{dia})] can become dominated by the background residual
rather than by PSF kernel residuals.

Whilst these effects are important for high surface brightness regions such as
the M31 bulge they are far less important for Galactic surveys that employ
difference imaging. Even towards the Galactic bulge the flux observed by
optical surveys is dominated by resolved or semi-resolved stars and therefore
any background residuals from systematic problems will be typically at a low
level even on difference images. However, it is feasible that photometric
corrections might benefit future near-infrared surveys of the Galactic bulge,
which can target the centre of the Galaxy where the unresolved stellar
background flux is considerably higher. Near-infrared array technology is also
still maturing and so is currently more prone to systematic
problems than optical CCDs.

The ISIS software also provides the option of running the OIS algorithm
independently within
sub-regions of an image.
In Figure~\ref{isis-bg}(b) we show the difference image result where ISIS
has split the image into $4 \times4$ sub-regions. All other DIA parameters
are the same as for Figure~\ref{isis-bg}(a). This time ISIS does a better job
over much of
the image area. However, within the inner bulge region, covering
around half the field area, noticeable background residuals are still evident,
especially at the square sub-region boundaries, which show sharp discontinuity.
One might be tempted to try to reduce the effect by further increasing the
number of sub-regions. However, we are fundamentally
limited by the size of the image stamps within which the
convolution kernel is determined, which in turn is limited by the scale of the
image PSF. For
the image in Figure~\ref{isis-bg}(b) each sub-region is only around 200 pixels
wide, compared to a stamp width of 30 pixels. Ideally the PSF and background
should be evaluated within several independent stamps across the sub-region in
order to facilitate a good solution for the PSF transformation. The sub-regions
also should be large enough to contain resolved or partially resolved stars, or
strong features such as dust lanes, so that the solution to equation~(\ref{dia})
is not dominated by noise. 

These limitations force us to consider treating the differential
background matching separately from PSF matching.

\begin{figure*}
\begin{center}
\includegraphics[width=\textwidth]{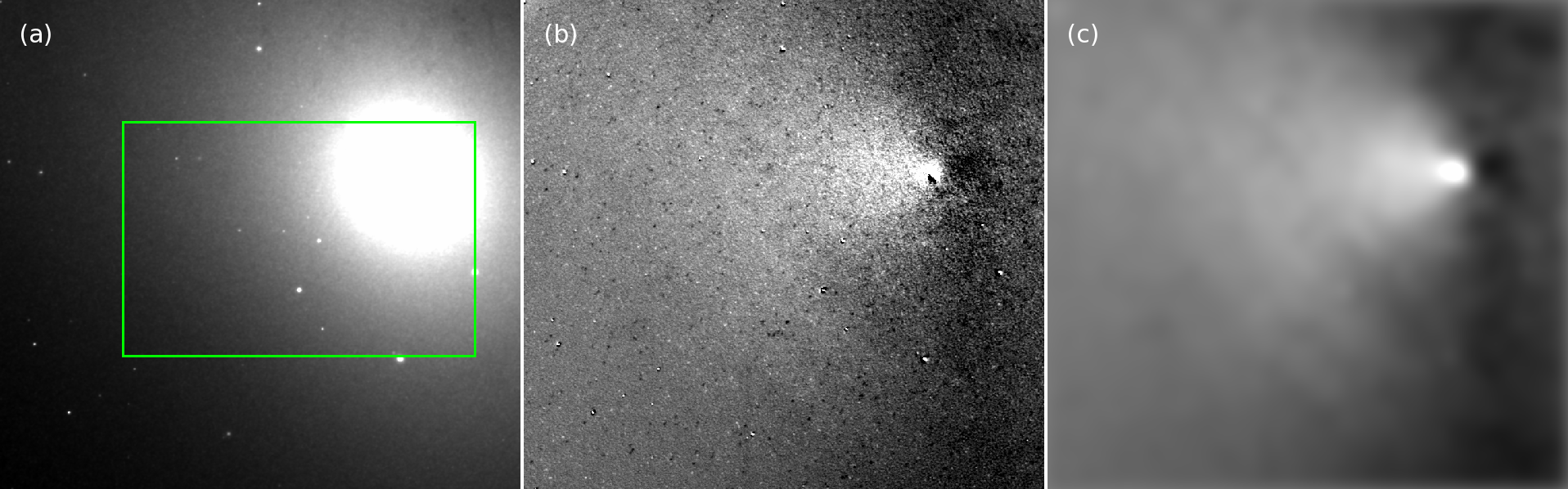}
\end{center}
\caption{ Matching the non-linear background residual. A target image
(a) is linearly matched to the reference
image [shown in
Figure~\ref{angfld}(a)] via a fit to the
pixels within the rectangular boundary, a region common to all exposures. The
resulting linearly aligned image is then directly subtracted
from the reference to reveal a residual
map (b). The map reveals a large scale non-linear variation in the residual
background level. This map is then Gaussian filtered to produce a smoothed
residual
background map (c). The contrast
level in (b) and (c) has been stretched by a factor 30 over that in (a) to
highlight the background residual, which has an amplitude that is typically only
about $1\%$ of the target image flux. The smoothed residual map
is added back on to the target image to produce the
final photometrically corrected target frame. The difference image resulting
from the corrected target frame is shown in Figure~\ref{isis-bg}(c).}
\label{bgmatch}
\end{figure*}

\section{A modified procedure for difference imaging} \label{moddia}

We have seen towards high surface brightness targets such as M31 systematic
effects such as scattered or reflected light can imprint complex
time-variable signatures upon the differential background
that the OIS algorithm cannot deal with adequately.
The OIS algorithm performs PSF and background flux matching
simultaneously, whilst prior to OIS these were treated separately
by various M31 variable photometry pipelines \citep{tom96,ans97}.
However, the OIS algorithm provides clear advantages over
earlier schemes due to its flexibility in modelling the PSF.
Ideally we would like to combine the best of the old and current approaches.

To this end we choose to separate out the photometric and PSF matching stages,
as in earlier approaches. Then
we run ISIS with differential background
matching effectively turned off by setting the
background polynomial to order zero
[i.e. setting $M_b = 0$ in Equation~(\ref{bgbasis})]. In this case the least
squares minimization of $D^2$ in Equation~(\ref{dia}) should 
be driven by the
quality of the PSF transformation kernel $K$ rather than potentially having to
trade between the quality of PSF and background matching.

\begin{figure}
\begin{center}
\includegraphics[width=0.48\textwidth]{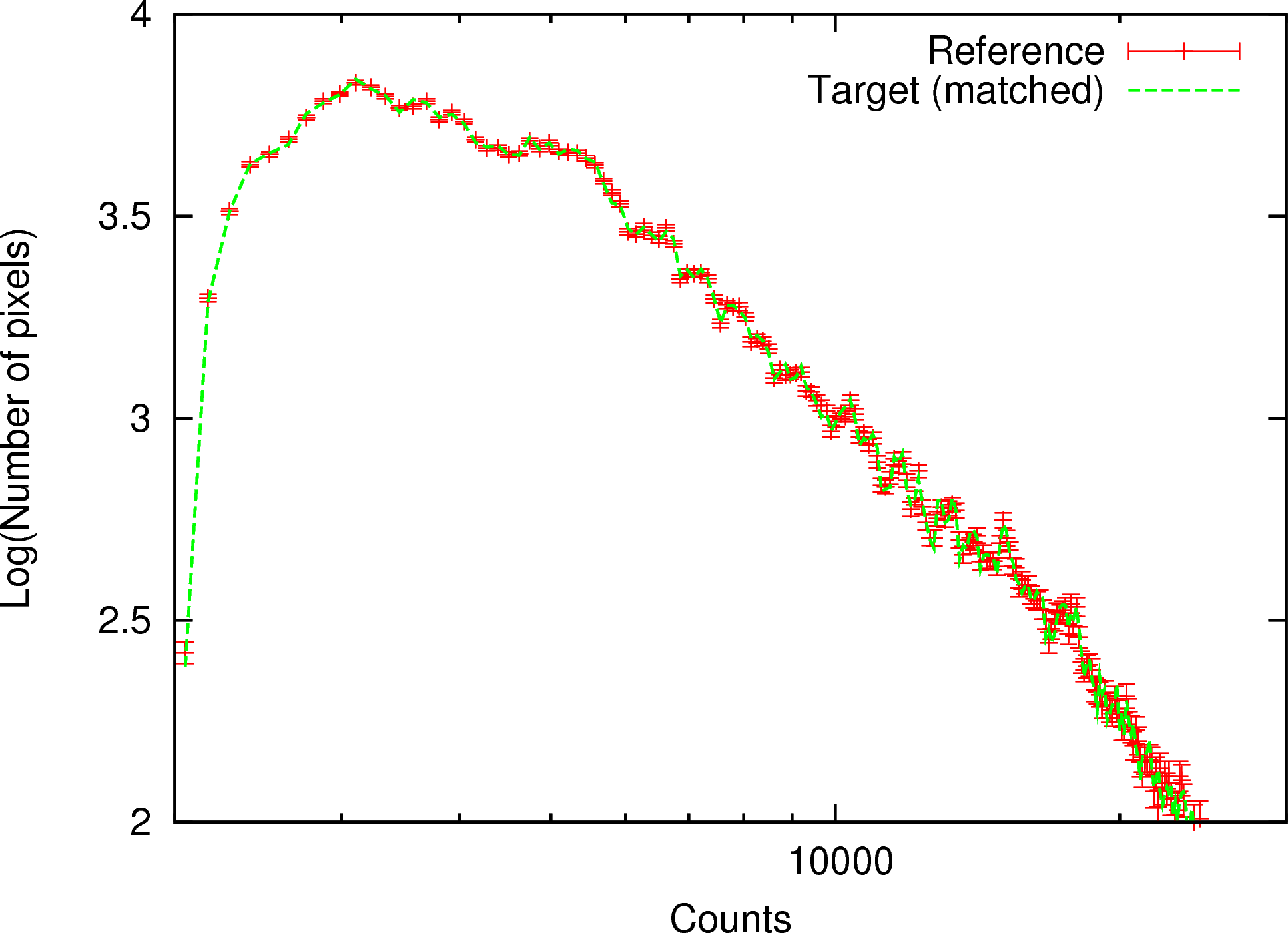}
\end{center}
\caption{Flux count
histograms of the reference image (points with errors)
and target image (dashed
line) after the target has been photometrically aligned to the reference. The
pseudo errors on the reference image pixel numbers assume an expected
Poisson-like variance in the pixel numbers between aligned images. After
alignment the behaviour of the target count histogram accurately mirrors that of
the reference image even down to
small scale deviations.}
\label{align}
\end{figure}

\begin{figure*}
\begin{centering}
\includegraphics[width=\textwidth]{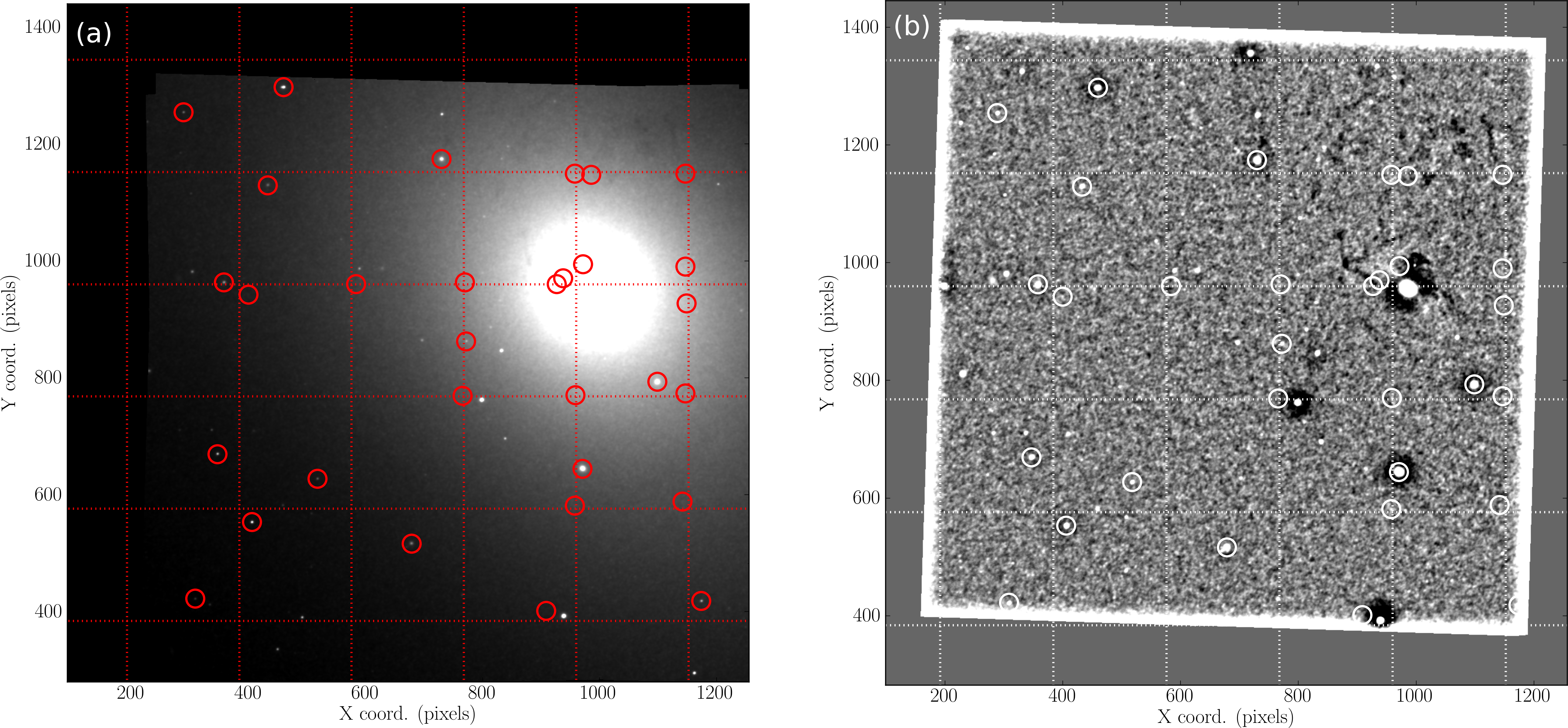}
\end{centering}
\caption{The reference image and target ``structure map'' aligned to
the same global pixel coordinates. (a) The reference image, showing the
locations of 31 objects
(shown circled - mostly foreground stars or features relating to dust lanes)
selected by ISIS to determine the
convolution kernel. One object is selected within each of the
square sub-regions
which form an
overlapping mesh indicated by the dotted lines. In (b) we show a
noise-normalised map  of residual structure in the target image. The
structure map is formed by
subtracting a smoothed version of the target and then dividing the result by the
square-root of the target flux. Foreground stars show up clearly on the
structure map, as do dust lanes close to the M31 core and in the upper right
of the image. The granular structure in the image is not noise but
partially resolved bright stars
and unresolved clumps of stars in M31. The smoothing aperture produces the dark
ringing effect evident
around some of the brighter foreground stars and the M31 core, as well as the
border around
the image perimeter.}
\label{stamps}
\end{figure*}

\begin{figure}
\begin{center}
\includegraphics[width=0.48\textwidth]{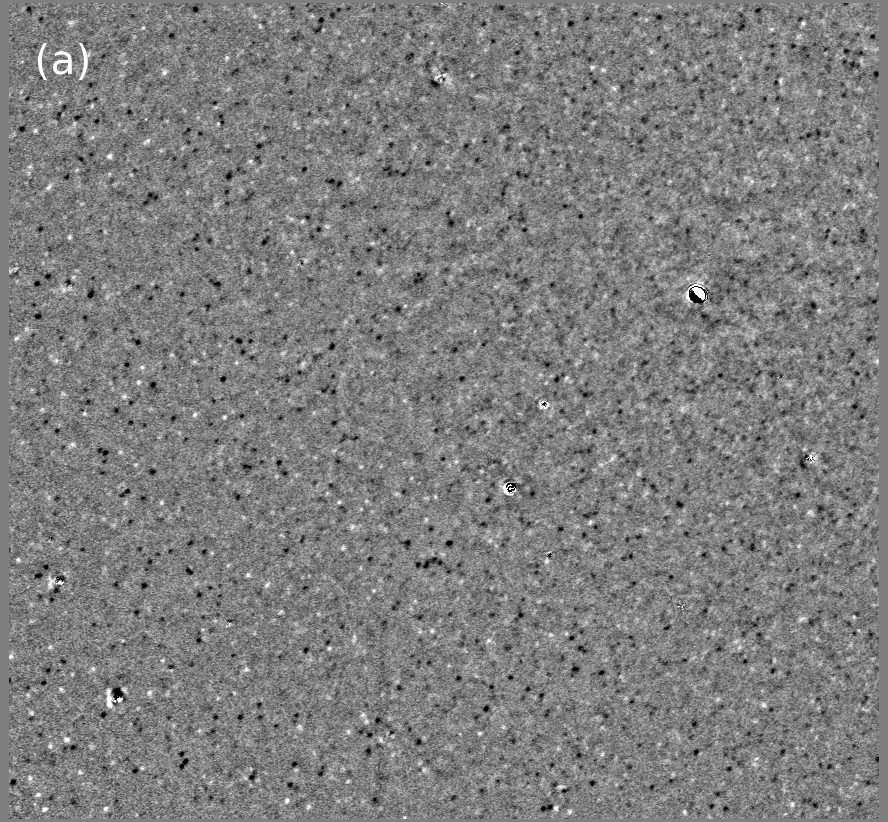}
\includegraphics[width=0.48\textwidth]{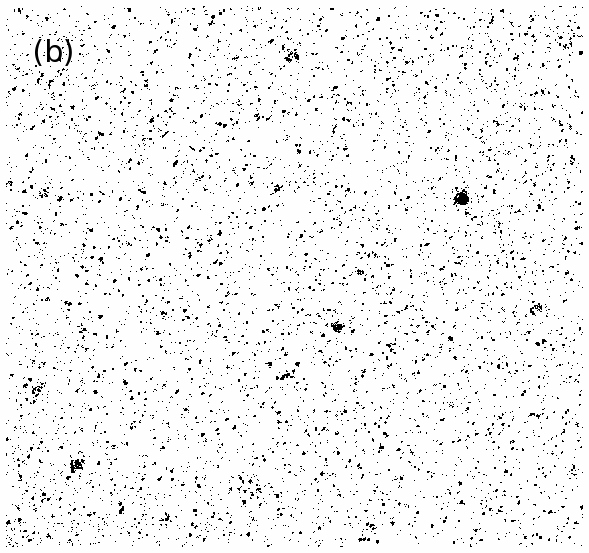}
\end{center}
\caption{(a) Grey-scale ``signal-to-noise'' map produced by dividing
the difference image
shown in Figure~\ref{isis-bg}(c) by the square root of the combined
reference and target image counts. Note how the increase in image noise towards
the M31 core suppresses the number of detectable variations there (compare the
contrast of sources close to the core with those further out) though some
are still evident to within $\sim 10''$ of the core. (b) A map of all the
pixels in (a) that exceed $2.5\,\sigma$. Isolated random noise pixels are
comparatively rare and most
belong to PSF-like clusters that are associated with genuinely variable
objects. These provide the excess seen in the tails of the noise histogram shown
in Figure~\ref{sigma-hist}.}
\label{sigma-map}
\end{figure}

\begin{figure}
\begin{center}
\includegraphics[width=0.48\textwidth]{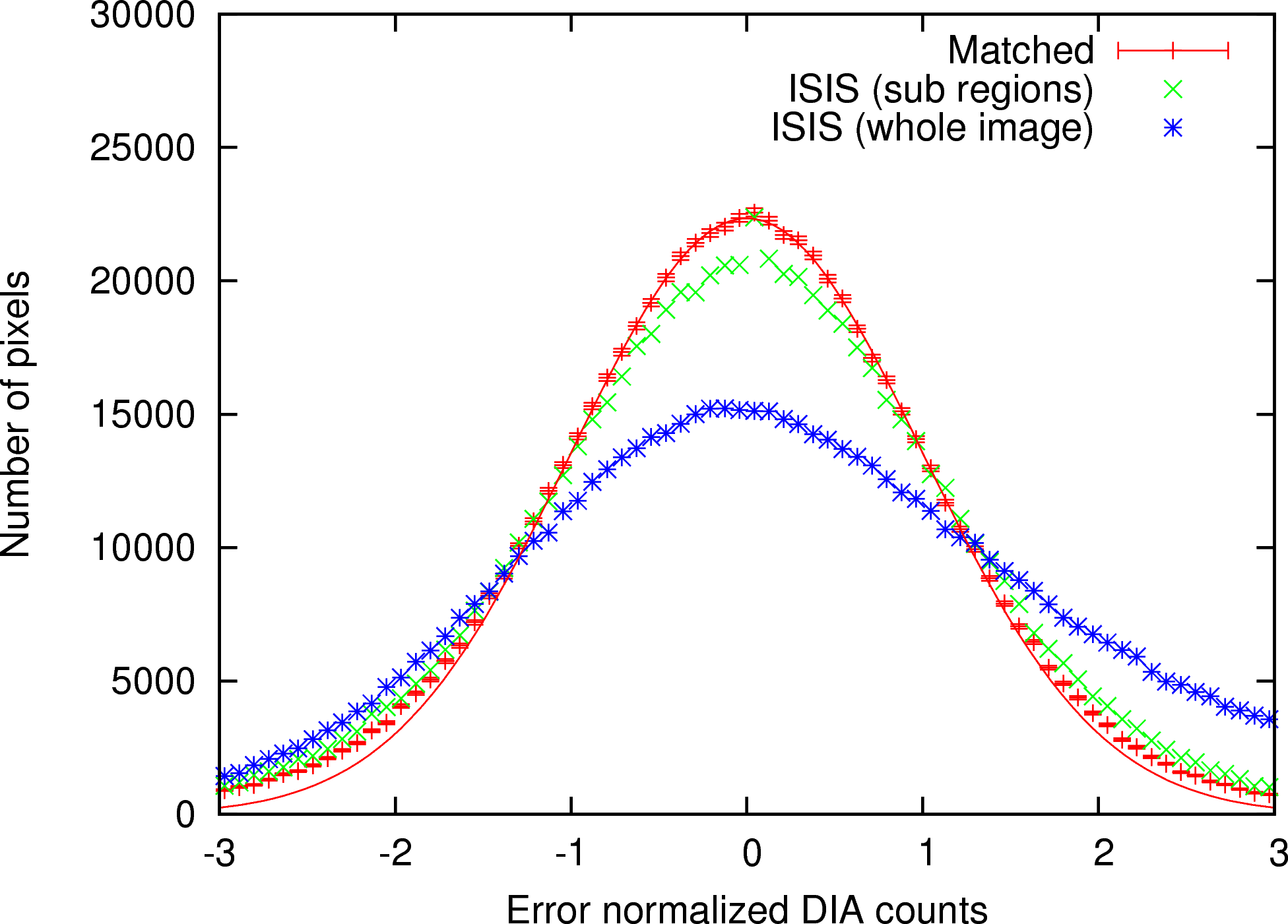}
\end{center}
\caption{Histogram of
the counts for the image shown in Figure~\ref{sigma-map}(a)
together with a unit-width Gaussian model. The amplitude of the Gaussian is
fitted to data points within $\pm 1 \,\sigma$. The excess from
pure Gaussian noise evident in
the tails of the histogram is predominately associated with genuinely variable
objects as shown in Figure~\ref{sigma-map}. Also shown are the equivalent
histograms for the difference images
shown in Figure~\ref{isis-bg}(a) and (b) where no photometric alignment is
performed prior to running ISIS.}
\label{sigma-hist}
\end{figure}

\subsection{Photometric alignment} \label{photscale}

We begin by determining a gross linear photometric scaling to match the image
flux. Due  to large telescope pointing errors in
the first two seasons of the Angstrom survey the sky area common to all
exposures represents less than half the available image area. We therefore
use the largest available rectangular region
common to all images, which covers around $30\%$ of the image area (see
Figure~\ref{bgmatch}) in order to
photometrically align our images.
We determine a gross linear
scaling and offset
between the reference $R$ and target $T$ ($R=a\times T+b$) using
the {\sc linmatch} task within \iraf. The
scaling $a$ and offset $b$ are then applied to $T$ to obtain a linearly
photometrically matched
image $T_L$.
After linear scaling we subtract $T_L$ from $R$ to obtain a residual image
$R-T_L$. Since the seeing is generally different between $R$ and $T_L$
the differential flux from genuinely variable sources is significantly smoothed
out by
differences in image PSF. A Gaussian filter is then passed over the
residual image to produce a smoothed residual
background image $B'$. The size of the Gaussian is set to $\sigma = 15$ pixels,
which is large enough to be insensitive to the presence of variable stars
but small enough to provide a good local estimate of the differential
background. Finally a background corrected target image $T' = T_L + B'$ is
obtained. Figure~\ref{bgmatch} illustrates how the differential background map
$B'$ is
produced from a target image which has already been linearly matched to the
reference frame.

Figure~\ref{align} shows the resulting pixel flux histograms after
photometric matching. Around the M31 bulge
the image flux is dominated almost everywhere by the smooth unresolved surface
brightness distribution, so the photometric calibration is insensitive to
seeing and we are able to obtain very good photometric alignment without the
need for PSF corrections.

The adjusted image $T'$ now
has
a background that is properly matched to $R$ so that ISIS can be
run without the inclusion of the $B$ term in Equation~(\ref{dia}). Note that
this matching is effective only against systematic effects which are additive
(e.g. scattered light). One might imagine that a poorly constructed flat field
could cause a non-uniform multiplicative variation across the image, in which
case the method described here would not adequately correct the flux.

\subsection{PSF matching and difference image results} \label{isis}

Before running ISIS on our photometrically aligned data we make checks on
whether there is enough information on the PSF within our images for ISIS to
make a reliable PSF kernel determination. Since the surface brightness of the M31
bulge is high the number of obvious resolved stars is relatively small.

ISIS determines the PSF kernel transformation not by using all the image
pixels but by sampling the image within a number of small regions (``stamps'')
dispersed over the image. The location of the stamps can be partly constrained
by the user specifying a grid of sub-regions. ISIS determines the
most suitable stamp location within each sub-region, usually centring the stamp
on a bright star. The quality of the PSF transformation is therefore partly
determined by the number of user-defined sub-regions and by the success with
which ISIS finds suitable stamps within each sub-region.

Figure~\ref{stamps} shows the reference image (a) and ``structure map'' of our
test target image (b) on which is over-plotted the sub-region grid adopted for
the tests in this paper. The structure map is produced by subtracting a
Gaussian smoothed version of the target image from the original target frame
and then dividing the result by the square root of the target pixel flux.
This removes the smooth background light component of the
M31 surface brightness, clearly revealing foreground Milky Way stars
as well as dust lanes, semi-resolved bright stars and stellar clumps
in M31 itself. The structure map clearly shows that there is a wealth of
structure over the whole image area which is potentially useful for PSF
determination.

The grid overlay in Figure~\ref{stamps}(a) and (b) is defined with respect to a
$1536\times 1536$ pixel area which defines a global pixel coordinate system onto
which all images have been aligned. (The large relative size of this area is
necessitated by large pointing offsets in some of the earlier exposures.) The
circles in Figure~\ref{stamps}(a) and (b) indicate the locations of the centre
of each of 31 stamps which ISIS selected for our test target frame. We set
the size of each stamp to $30\times 30$ pixels, so that collectively the stamps
comprise about $3\%$ of the available image area. More often than not the stamps
are centred on bright foreground stars, however it is clear from
Figure~\ref{stamps} that a number of the stamps are not centred on any
obvious star. We find that the steep surface brightness gradient is a strong
determinant as to whether the ISIS stamp finding algorithm locates a bright star
or not. 

In order to determine the PSF kernel reliably, the number of stamps should be at
least $3(1+M_k)(2+M_k)/2$ \citep{ala98,ala00}, where $M_k$ is the user-specified
polynomial order allowed for the spatial variation of the kernel (see
Equation~\ref{basis}). For our
data we find $M_k = 1$ sufficient to characterise spatial PSF kernel variations,
so that we require a minimum of 9 good stamps. In fact, typically the ISIS stamp
finding algorithm results in $\sim 15$ stamps centred on genuine bright
foreground
stars, more than sufficient
to compute a reliable kernel. We have also checked
that using an improved star finding algorithm, in which all the stamps are
centred on obvious stars, makes little difference to the overall difference
image result. In fact, whilst the bright foreground stars are preferred for PSF
determination, the granular structure of partially resolved M31 stars evident in
Figure~\ref{stamps}(b) means that just about any position on the image will
provide some useful information for the PSF kernel determination. We therefore
proceed with the default stamp finding algorithm which is contained in the ISIS
package.

The result of running ISIS on a target image that is now photometrically
aligned to the reference frame is shown in Figure~\ref{isis-bg}(c). All ISIS
configuration parameters are
set the same as for the previous tests in Figure~\ref{isis-bg}(a) and (b) except
we 
now enforce a zeroth-order polynomial model for the differential background. The
resulting
difference image is clearly superior to the previous examples. There is now no
noticeable background residual anywhere on the image. Equally important is the
fact that the PSF transformation is clearly much improved too. This is evident
when examining the residuals associated with the few bright foreground stars
visible within the Angstrom field (compare with the reference image shown in
Figure~\ref{angfld}). These foreground stars leave relatively strong residuals
in the
difference images shown in Figure~\ref{isis-bg}(a) and (b) but they are
much reduced in Figure~\ref{isis-bg}(c) (see also
the zoomed inset panels). This indicates that ISIS has been able to
find a better PSF solution as a result of minimizing Equation~(\ref{dia})
for photometrically aligned images.

Figure~\ref{sigma-map}(a) shows a map of the signal to noise ($S/N$) ratio.
The signal-to-noise map is defined by
   \begin{equation}
 S/N = \frac{\sqrt{gMN}(R-T)}{\sqrt{MT+NR}}, \label{sn}
\end{equation}
where $M$ is the number of individual exposures comprising the target frame
stack $T$, $N$ is the number of individual exposures comprising the reference
stack $R$ and $g$ is the CCD gain. The grey-scale in Figure~\ref{sigma-map}(a)
is
calibrated to show pixels with $|S/N|
>
10$ as either white or black depending upon whether the difference flux is
positive or negative. The main difference in the appearance of this $S/N$ image
with the corresponding difference image shown in Figure~\ref{isis-bg}(c) is
that near the M31 core the $S/N$ image shows much less variation, indicating
that many of the flux deviations on the difference image near the M31 core are
just due to the noise arising from the very high surface brightness level.
Figure~\ref{sigma-hist} shows the resulting pixel histogram
of $S/N$ values for the photometrically aligned target frame and for the
unaligned cases shown in Figure~\ref{isis-bg}(a) and (b). These histograms
demonstrate a clear improvement both in the symmetry and width of the noise
distribution as a result of carrying out separate photometric alignment.
Over-plotted on the noise histogram of the aligned data is a
unit-width Gaussian model, where the amplitude has been fitted within $\pm 1
\, \sigma$. The histogram shows that within this region the noise is well
approximated
as Gaussian. Towards the tails of the distribution there is an excess over and
above Gaussian noise. However, as shown in Figure~\ref{sigma-map}(a) and (b),
this excess is associated with
genuinely
varying objects which make up a
non-negligible fraction of the image pixels.
Figures~\ref{sigma-map} and \ref{sigma-hist} therefore
suggest that photometrically aligned
images can allow high quality difference imaging even to within a few
arcseconds of the M31 bulge.

\subsection{The effect of variables} \label{vartest}

The high crowding levels of genuinely variable sources evident in
Figure~\ref{isis-bg} potentially poses a problem for the OIS
algorithm. The minimization in Equation~(\ref{dia}) would be expected
to correspond to the
optimal difference image only in the limit that, in the absence of seeing and
background variations, most of the image flux does not
vary with time. In theory
it could be the
case that the relative phases and amplitudes of variables within a stamp at a
given
epoch could pose a significant source of time variability in the average
difference flux within the stamp. If so this could also affect the residual
background model determination. 

In order to suppress the effects of genuine
variations on the computed kernel ISIS uses sigma clipping both of pixel values
within stamps and of the distribution of stamps themselves. However, for the
M31 bulge, where a relatively high fraction of pixels change due to
astrophysical variation, sigma clipping can lead to a significant fraction of
the image area being excluded, particularly around the M31 core.

To test the potential impact of the high sky density of variables we perform
a double-pass iterative minimization of Equation~(\ref{dia}) that corrects the
target image for the presence of variable sources. On the
first pass ISIS is run in the normal way. Then, all pixel fluxes on the
resulting difference image above a specified absolute flux threshold are added
back on to the target frame to form a modified target image in which
intrinsically variable sources are largely removed (i.e. matched to their flux
values on the reference image). A second pass of ISIS is
then performed on this modified target frame. The resulting kernel solution for
the modified frame is then applied to the reference image and the convolved
reference is subtracted from the original (unmodified) target frame to create
the final difference image. This method ensures that the final difference
image is
largely insensitive to the presence of variable stars and that as much of the
image area as possible is used to compute the kernel.

Reassuringly, we find virtually identical results are recovered using either
a standard ISIS run or the double-pass method described above, showing that
ISIS is capable of finding an optimal kernel solution
even for high crowding densities of variable stars. 

\section{Example light-curves} \label{examplelc}

\begin{figure*}
\begin{centering}
\includegraphics[width=0.48\textwidth]{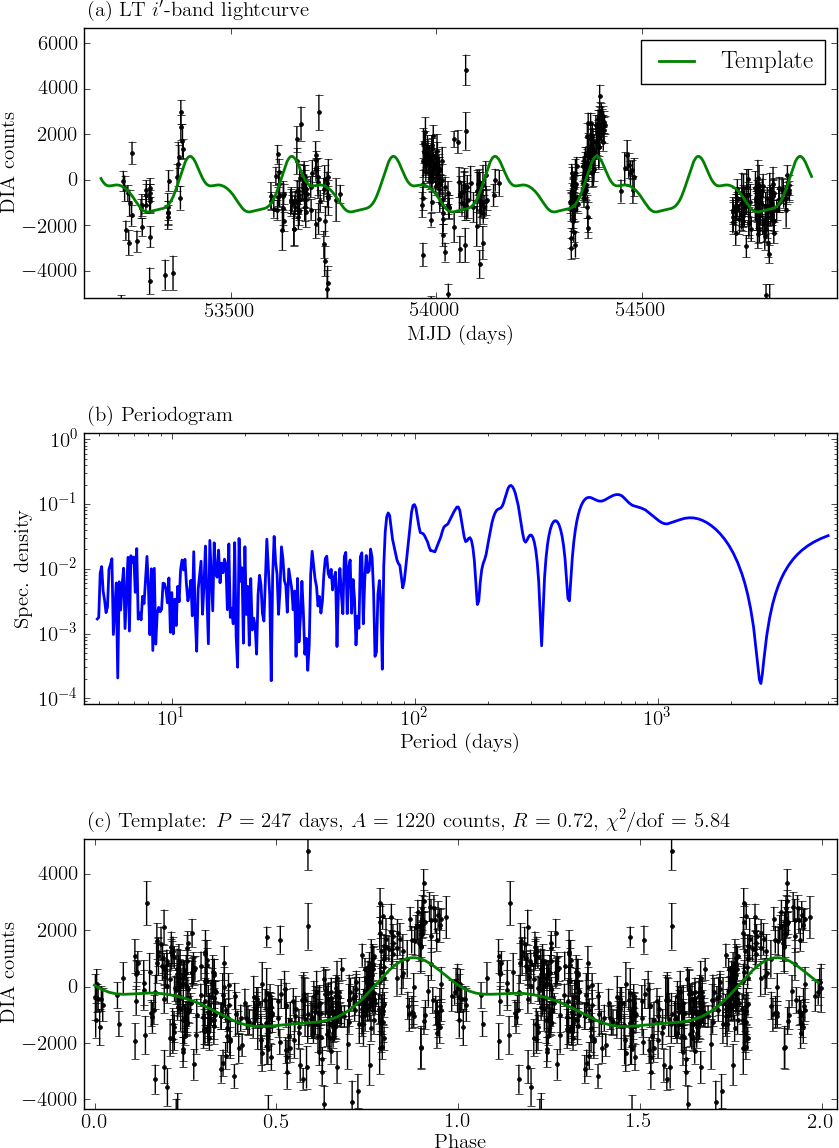}
\includegraphics[width=0.48\textwidth]{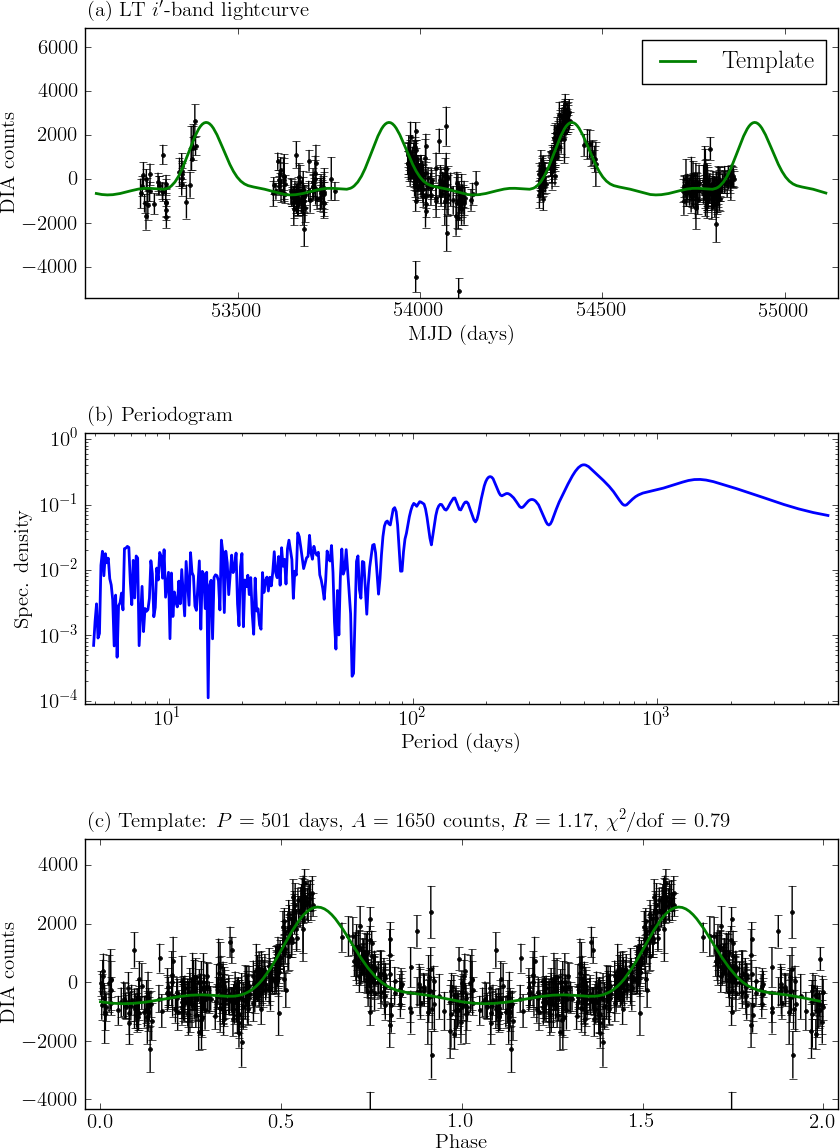}
\end{centering}
\caption{A five-season LT $i'$-band light-curve of a
periodic
variable star located $2\farcm 4$ from the M31 core. Results using ISIS
applied to photometrically unaligned images are shown in the left-hand column
whilst those from aligned images are shown in the right-hand column. In each
case panel (a) shows the light-curve, panel (b) shows the spectral density
computed using the Generalised Lomb Scargle method of \protect\cite{zech09} 
and panel (c) shows the result of folding the light-curve in (a) by the period
corresponding to the maximum in the spectral density shown in (b). A smoothed
light-curve template constructed using radial basis function interpolation (see
main text) of the folded light-curve is indicated by the smooth curve in (a) and
(c). In (c) the folding
period $P$, the variable half amplitude $A$, the asymmetry ratio $R$ and the
reduced $\chi^2$ with respect to the template curve (excluding the 6 worst
outlying points) are also given. The
asymmetry ratio $R$ is the ratio of the rise (minimum up to peak) and fall (peak
back to minimum) time-scales, as determined from the smooth template. Note how in
the unaligned case the systematic scatter has resulted in an incorrect period
determination.}
\label{lc-comp}
\end{figure*}

\begin{figure*}
\begin{centering}
\includegraphics[width=0.47\textwidth]{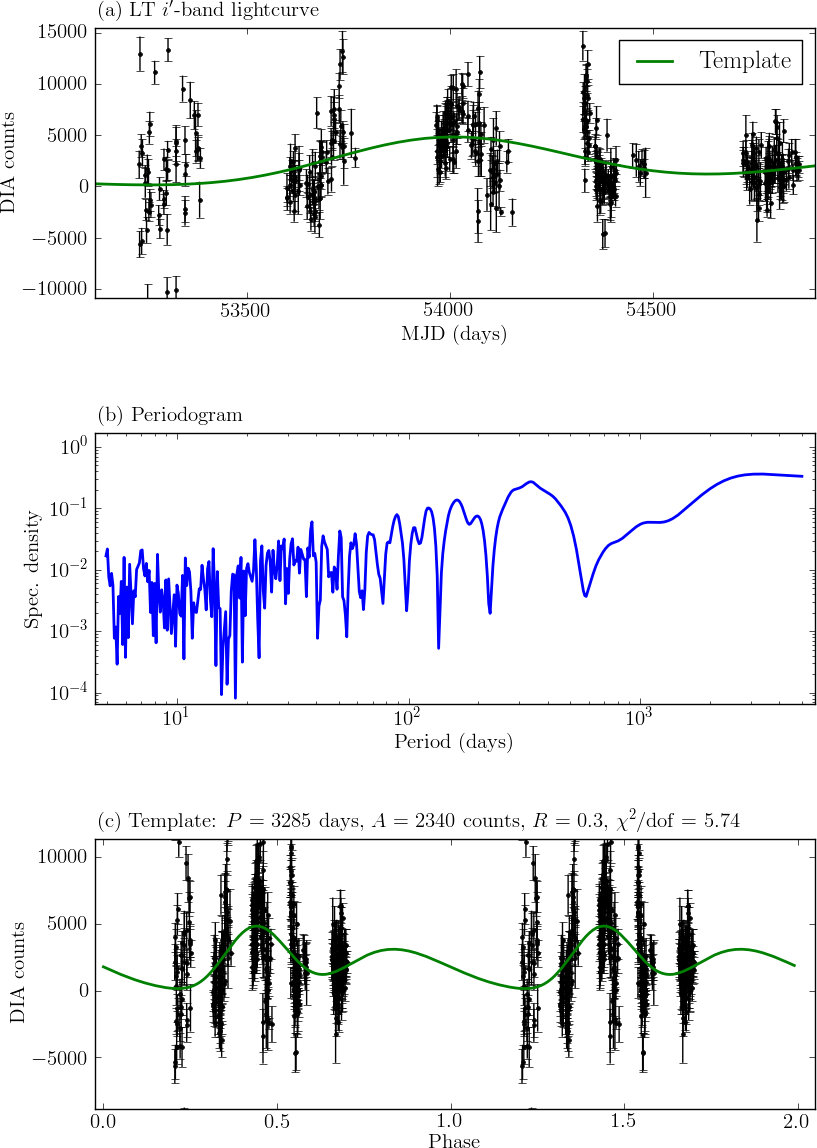}
\includegraphics[width=0.49\textwidth]{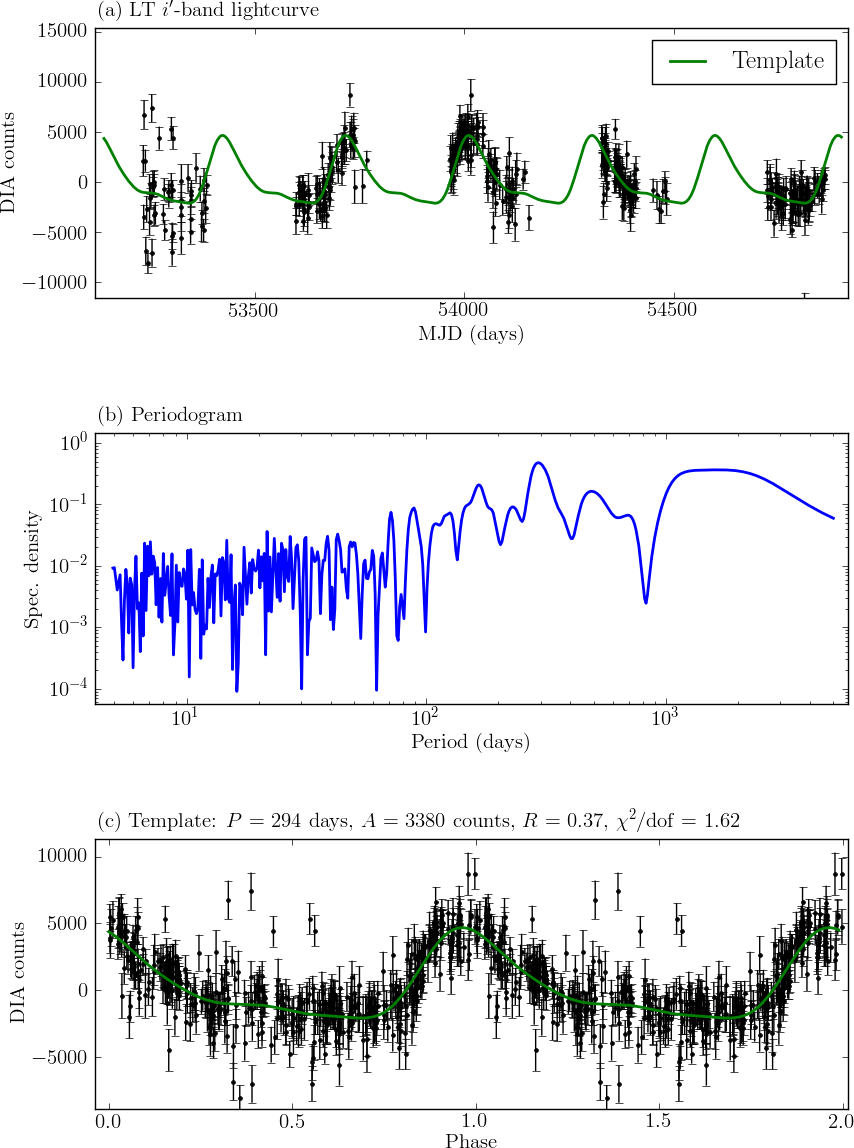}
\end{centering}
\caption{As for Figure~\ref{lc-comp} but this time for a variable star located
just $19''$ from the M31 core. In the unaligned case displayed in the
left-hand column the periodogram fails to identify the correct period and
instead the peak spectral density reflects merely the season-to-season
variation in average flux.}
\label{lc-comp2}
\end{figure*}

In Figures~\ref{lc-comp} and \ref{lc-comp2} we show two periodic variable star
light-curves from the
Angstrom Project data processed with and without the photometric alignment
stage described in Section~\ref{photscale}. Figure~\ref{lc-comp} shows the
light-curve of a variable with a period of 501~days located $2 \farcm 4$ from
the M31
core monitored over five observing seasons. The right-hand and left-hand
columns show the results  with and without photometric alignment,
respectively. Panel (a) shows the light-curve itself, panel (b) is the computed
periodogram, and panel (c) is the result of folding the light-curve
with the period corresponding to the peak in the periodogram.

Looking at panel (a) in the right-hand column
of Figure~\ref{lc-comp} we see that the light-curve derived from the
photometrically aligned images appears much more coherent and less
scattered than that derived from unaligned data in the left-hand panel. Panel
(b) shows the computed periodogram for periods from 5 to 5000 days calculated
using the Generalised Lomb Scargle method of \cite{zech09}. This generalised
method extends the
standard Lomb Scargle approach by taking account of errors on the flux and
and not assuming a zero mean flux, making the periodogram computation more
reliable when there are large time gaps in the data (such as for seasonal
surveys like ours). The period corresponding to the peak in the spectral
density is used to fold the light-curve data in panel (c).
The folded data in the right-hand column clearly exhibits well defined
periodic behaviour. 

In order to provide a guide on the performance of the photometry we compute the
reduced $\chi^2$ with respect to model templates of the light-curve
behaviour.
We assume the underlying behaviour of the folded
data to be smooth on temporal scales much larger than the time resolution of
the folded data. This may not always be the case so the resulting $\chi^2$
assessment should be conservative in the sense of not being an over-optimistic
indicator of the true performance. We construct a template for each light-curve
using the {\tt
SciPy}\footnote{\url{http://www.scipy.org/}} implementation of the multiquadric
radial basis function (RBF) approximation, which is a generalisation to the
case of noisy data of multiquadric RBF interpolation \citep{har71}. The RBF
approximation is controlled by two parameters: a
shape parameter ($\varepsilon$) and a smoothness parameter ($S$).
The
shape parameter controls the way in which the
interpolation function weighting varies with Euclidean
distance (norm) from the data points, whilst the smoothness parameter controls
the level
of smoothing over scatter in the data. Setting $S = 0$ would be equivalent
to strict interpolation, where the template is forced through every data point
(clearly inappropriate for noisy data). We fix these parameters throughout at
$\varepsilon = 0.25$ and $S = 0.05$ which, for adequately sampled
light-curves, provide a good description of trends in light-curve behaviour
without appearing to over-fit the data; light-curve ``structure''
below $\sim 1/4$ of a phase is typically unresolved. For the
computation of the multiquadric 
RBF approximation it is desirable to first rescale the flux axis so that both
the flux and phase axes have similar numeric scales (of order unity). This
ensures that the interpolation weighting is sensitive to displacements along
both axes.

Looking at the left-hand panels of Figure~\ref{lc-comp} we see that the
systematic scatter in the light-curve derived from photometrically
unaligned data results in a significantly distorted periodogram which
peaks at the wrong period. Since for our DIA testing we have not
attempted to exclude poor quality images the light-curves all contain a
small number of obvious outliers. When computing the reduced $\chi^2$
with respect to the
RBF template we therefore exclude the 6 worst outlying points (representing
$1.25\%$ of the full dataset) in order to obtain a fair measure of how well the
template
represents the bulk of the data. The reduced $\chi^2$ value reported in panel
(c) is quite
acceptable for the aligned data but clearly poor for the unaligned
folded dataset.

Figure~\ref{lc-comp2} shows a similar situation for a variable star with
a period of 294~days located just 19~arcsec from the M31 core,
corresponding to a projected distance from the core of just 72~pc for an M31
distance of
780~kpc. In the unaligned case the peak in the periodogram is
completely spurious, corresponding to the
seasonal variation in average flux due to different parts of the variable phase
being
sampled in different observing seasons. Once again the reduced $\chi^2$ for
the RBF template is quite acceptable for the light-curve derived from
aligned data but poor for the unaligned dataset. The examples of
Figures~\ref{lc-comp} and \ref{lc-comp2} show that careful photometric
alignment is essential for the accurate characterisation of variable signals.

It is noticeable that the first
season of Angstrom data shows good consistency with subsequent seasons for the
variable star shown in the right-hand column of Figure~\ref{lc-comp}, which is
 located $2 \farcm 4$ from the M31 core (about a half of the LT field width
away). However the light-curve of the variable close to the M31 core shown in
the right-hand column of Figure~\ref{lc-comp2} is much more erratic during the
first season than in
subsequent seasons. Visual inspection of the images shows that the majority of
images taken during the LT commissioning season in 2004 show evidence of
scattered or reflected light close to the core. Improvements in light baffling
made to the LT in early 2005 substantially improved the quality of later
data.

\begin{figure*}
\begin{centering}
\includegraphics[width=0.49\textwidth]{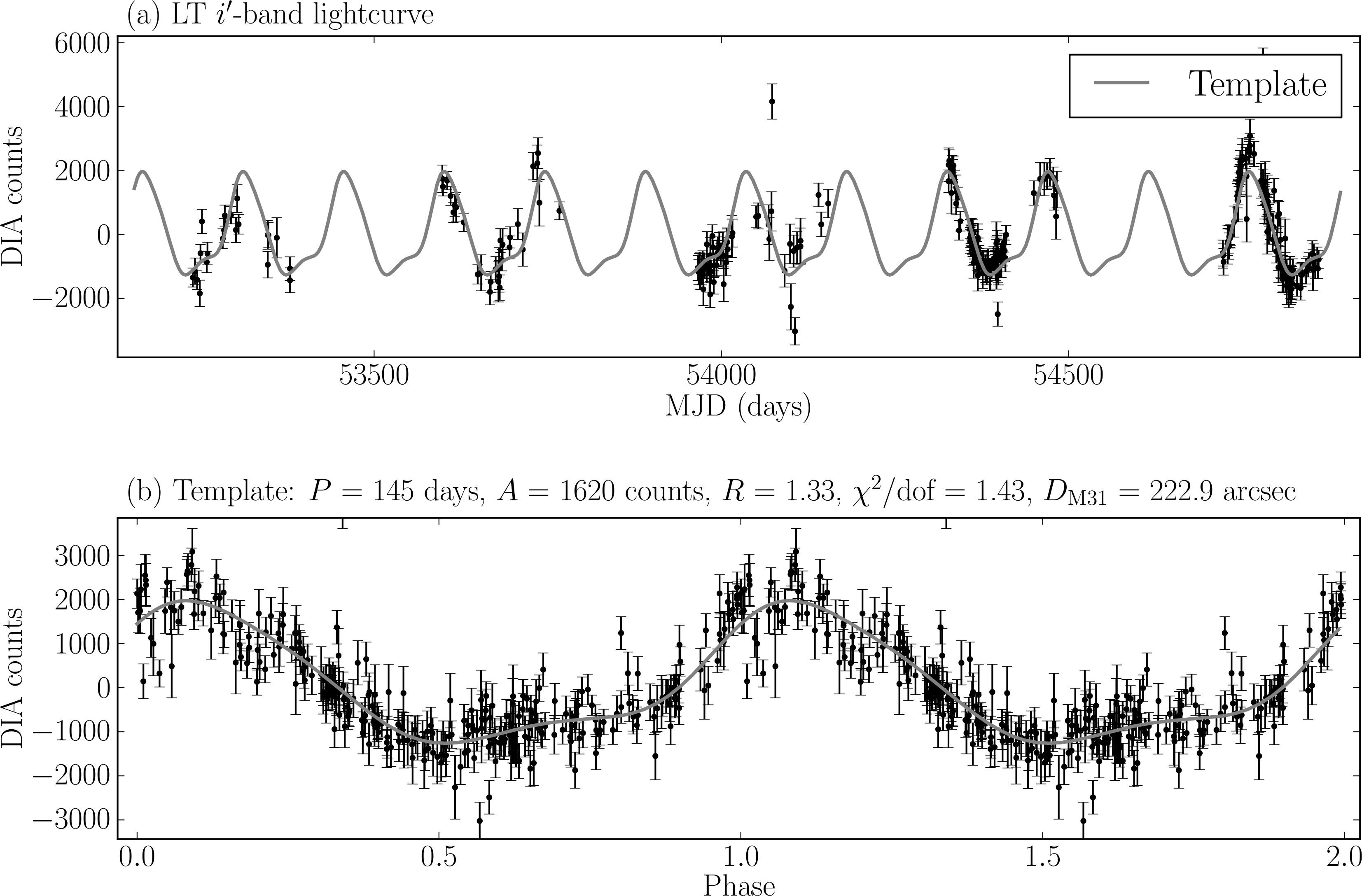}
\includegraphics[width=0.49\textwidth]{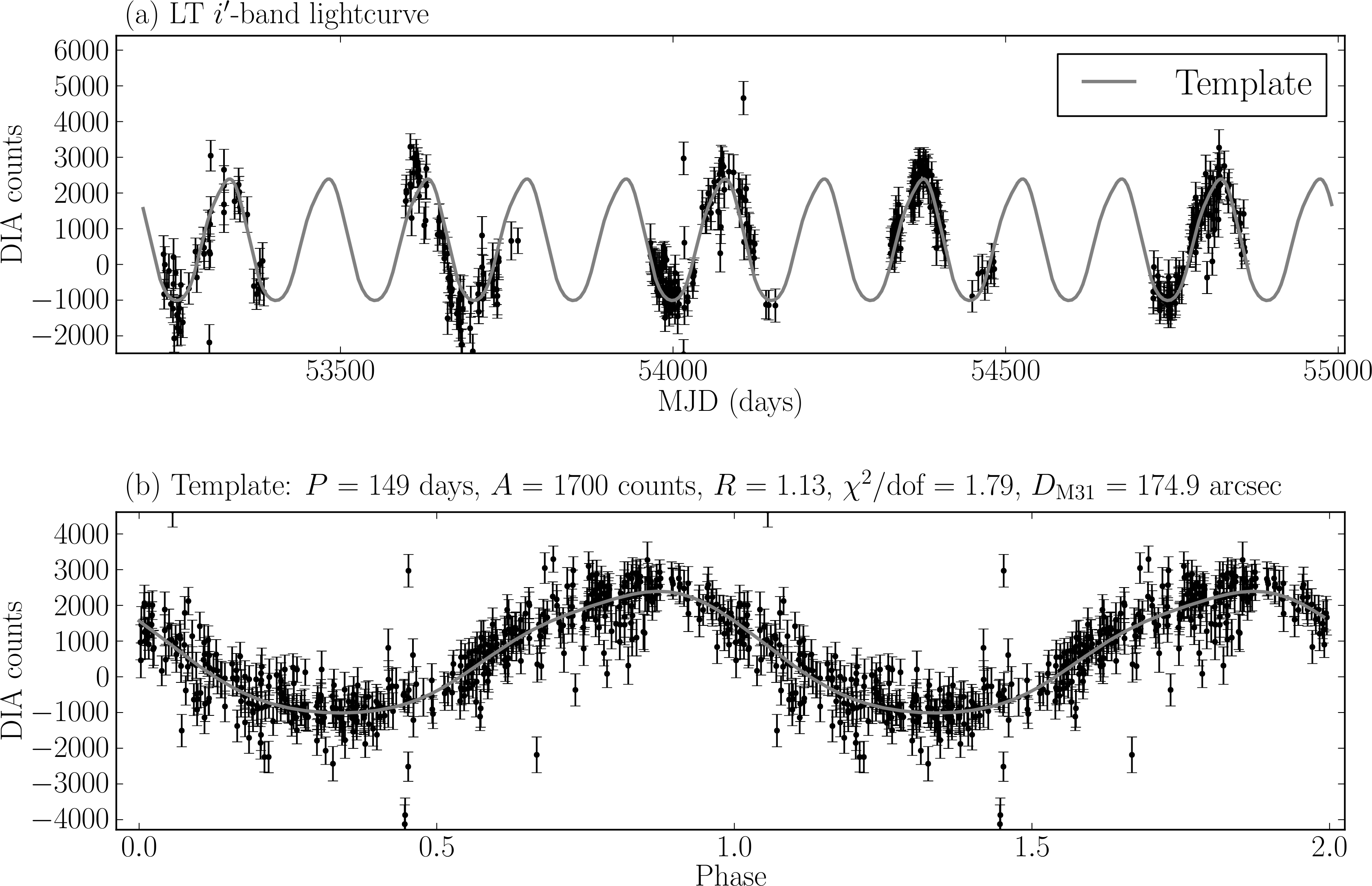}
\includegraphics[width=0.49\textwidth]{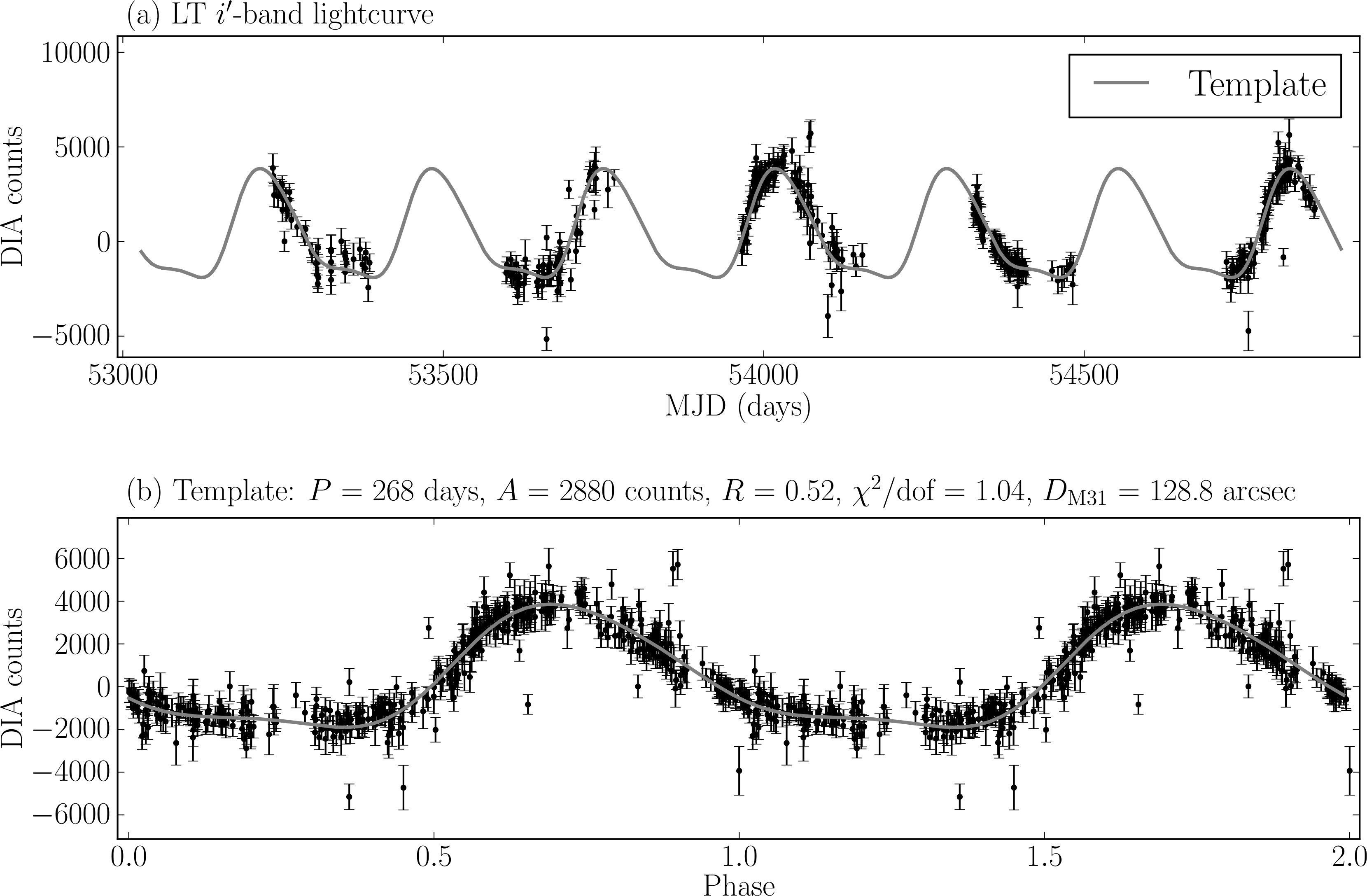}
\includegraphics[width=0.49\textwidth]{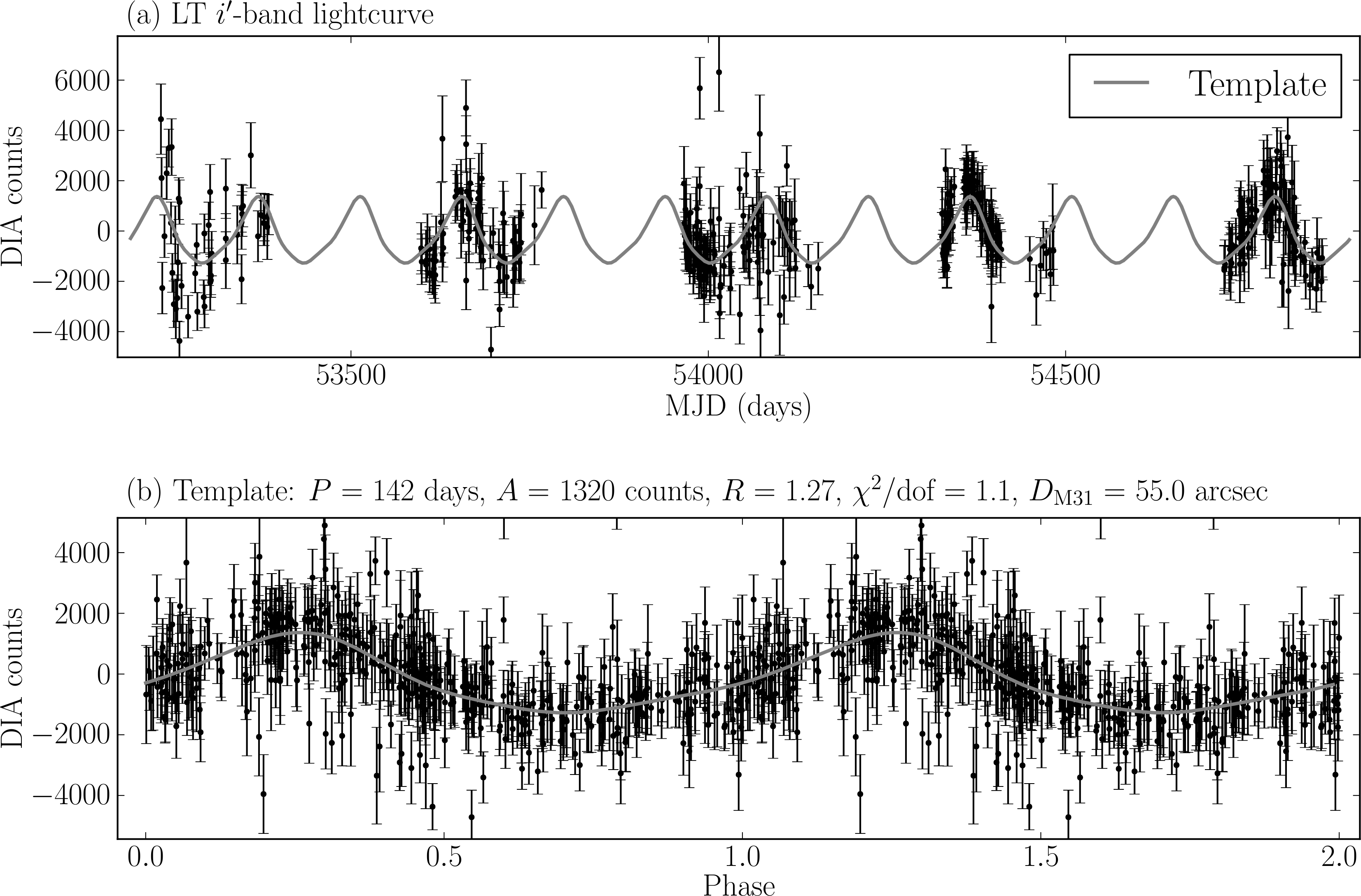}
\includegraphics[width=0.49\textwidth]{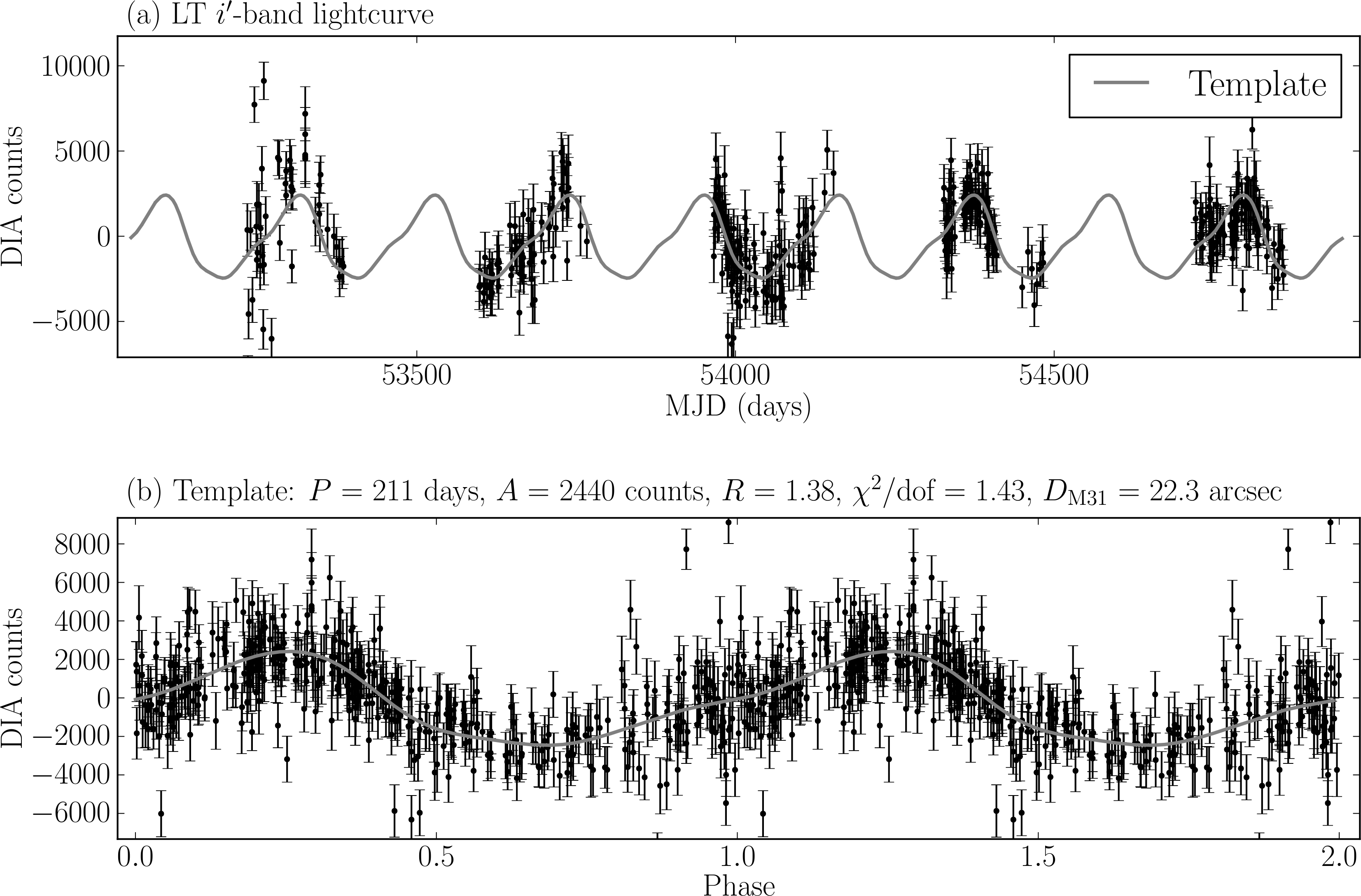}
\includegraphics[width=0.49\textwidth]{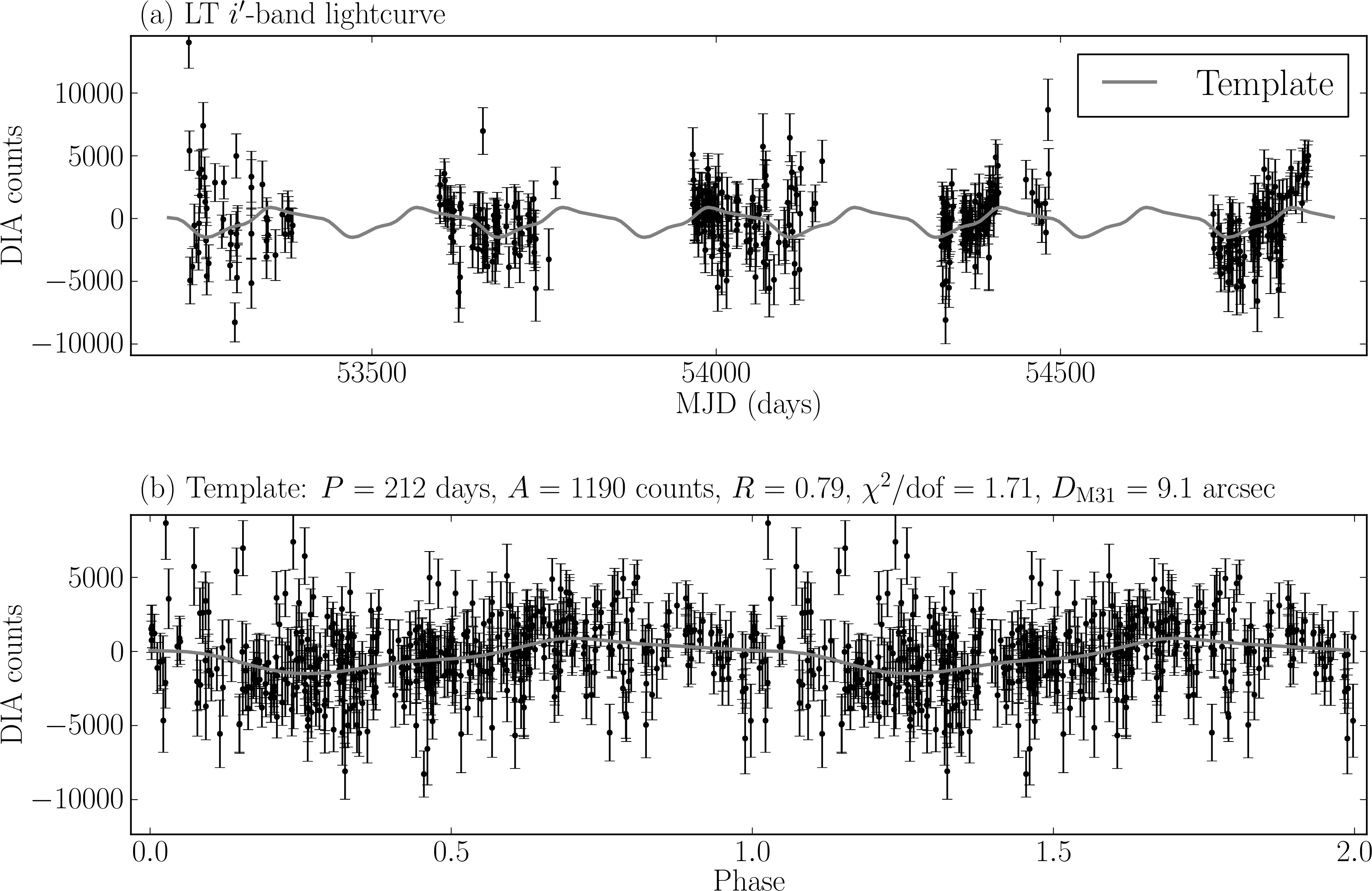}
\end{centering}
\caption{Corrected light-curves for a sample of six periodic variable stars over
a range
of distance from the M31 core. The five-season light-curves are shown in the
panels labelled (a) together with the smooth template determined from the
folded light-curve. The folded data is shown in panel (b) below each
light-curve.
The statistics displayed above the folded data is as for Figures~\ref{lc-comp}
and \ref{lc-comp2}, with the addition of $D_{\rm M31}$ which is the angular
distance of the object from the M31 core in arcseconds.}
\label{examples}
\end{figure*}

Figure~\ref{examples} shows six more examples of detected variable stars
at angular distances from the M31 core ranging from 223~arcsec (near the edge of
our reference field) down to just 9~arcsec. They illustrate good
season-to-season consistency in the corrected photometry, as evidenced by the
smooth
continuity of the folded light-curve, and also reasonable characterisation
of the noise, as indicated by the size of the reduced $\chi^2$ values with
respect to the RBF template. 

The variable in Figure~\ref{examples} located at 9~arcsec from the core has
a predictably very noisy light-curve and is at the threshold of detectability;
whilst it displays clear periodic behaviour the structure of the variation
is not well resolved due to noise. We therefore determine that the technique
works well at least beyond $\sim 10$~arcsec from the M31 core.
As Figure~\ref{sigma-map} shows, there are relatively few variable objects
that are detectable above the noise so close to the M31 core but the examples in
Figures~\ref{lc-comp2} and \ref{examples} demonstrate that, where they are
detectable, we are
usually able to obtain reliable photometry even in this extreme high-background
regime.

\section{Discussion} \label{discuss}

In this paper we have shown that optimal difference imaging in regions of very
high background levels, such as the bulge regions of galaxies, can be severely
compromised due to the presence of systematics such as internal reflections,
scattered light,  or fringe
effects. In these
cases difference images 
created using the Optimal Image Subtraction (OIS) algorithm \citep{ala98,ala00},
which is the most widely used difference image algorithm, may exhibit large
amplitude background residuals that make reliable relative photometry
difficult if not impossible.
Fortunately, we have shown that OIS is able to give very
good results provided the images to be differenced are first photometrically
aligned prior to difference imaging. 

Using the photometric alignment procedure described in this paper we find we can
produce difference images of the M31 bulge that are close to photon noise
limited. Not only does it
minimize or eliminate large amplitude background residuals but it also
noticeably improves
the quality of the PSF kernel transformation. We have tested the modified
pipeline by producing several examples of periodic variable light-curves from
the Angstrom Project survey of the M31 bulge. The results allow characterisation
of periodic signals even to within $\sim 10$ arcseconds of the M31 nucleus.

The problem we highlight is specific to targets in which the background
brightness is
high and subject to systematic variations which have
complex spatial signatures. The OIS method is well known to cope
admirably for stellar fields within our Galaxy, as imaged by current optical
surveys, where the image
flux is dominated by resolved or semi-resolved stars rather than by the diffuse
background light. However, future near-infrared time-domain surveys of the
Galactic Centre could also conceivably benefit from a separation of the
photometric and kernel matching stages.

\section*{Acknowledgements}

It is a pleasure to thank the referee, Przemyslaw Wozniak, for many useful
suggestions which helped to improve this paper. JPD was supported by a PhD
studentship from the UK Science and Technology
Facilities
Council (STFC). This work was supported by the
research grant of the Chungbuk National 
University in 2009.
The Liverpool Telescope is operated on the island of La
Palma by Liverpool John Moores University in the Spanish Observatorio
del Roque de los Muchachos of the Instituto de Astrofisica de Canarias
with financial support from STFC.

\end{document}